\documentclass[pre,twocolumn,aps,10pt,superscriptaddress,notitlepage,nofootinbib]{revtex4-1}
\usepackage{epsfig,bm,amsmath,amssymb,amsfonts,graphicx,color,calc,epstopdf,stmaryrd}

\newcommand{\beq}{\begin{equation}}
\newcommand{\eeq}{\end{equation}}
\newcommand{\ba}{\begin{eqnarray}}
\newcommand{\ea}{\end{eqnarray}}
\newcommand{\thav}[1]{\left<#1\right>}

\newcommand{\B}[1]{{\bm{#1}}}

\renewcommand{\H}{\mathcal{H}}
\newcommand{\A}{\mathcal{A}}
\renewcommand{\B}{\mathcal{B}}
\newcommand{\C}{\mathcal{C}}

\newcommand{\Var}{\mathrm{Var}}
\newcommand{\Cov}{\mathrm{Cov}}
\newcommand{\Exp}{\mathrm{E}}

\begin{document}

\title{Breakdown of Nonlinear Elasticity in Amorphous Solids at Finite Temperatures}

 \author{Itamar Procaccia}
 \affiliation{Department of Chemical Physics, the Weizmann Institute of Science, Rehovot 76100, Israel}

 \author{Corrado Rainone}
 \affiliation{Department of Chemical Physics, the Weizmann Institute of Science, Rehovot 76100, Israel}

\author{Carmel A.B.Z. Shor}
\affiliation{Department of Chemical Physics, the Weizmann Institute of Science, Rehovot 76100, Israel}

\author{Murari Singh}
\affiliation{Department of Chemical Physics, the Weizmann Institute of Science, Rehovot 76100, Israel}

\begin{abstract}
It is known by now \cite{HentschelKarmakar11} that amorphous solids at
zero temperature do not possess a nonlinear
elasticity theory: besides the shear modulus which exists, all the higher order coefficients do not
exist in the thermodynamic limit. Here we show that the same phenomenon
persists up to temperatures comparable to the glass transition. The zero temperature mechanism due to
the prevalence of dangerous plastic modes of the Hessian matrix is replaced by anomalous stress
fluctuations that lead to the divergence of the variances of the higher order elastic coefficients.
The conclusion is that in amorphous solids elasticity can never be decoupled from plasticity: the
nonlinear response is very substantially plastic.
\end{abstract}

\maketitle

{\bf Introduction}: By cooling glass forming liquids below their glass transition temperature one forms amorphous solids.  They are solid because particles are not free to move ergodically, but rather can only vibrate around equilibrium positions. They are amorphous because, differently from crystals, those positions possess no long-range periodicity. As a result of this, a glass sample is always unique: while the structure of a crystalline solid is always realized in the same manner (barring local defects), the amorphous structure of a glass is randomly selected~\cite{Dy06,CavagnaLiq}. So, even if an ensemble of glasses is prepared with a perfectly reproducible protocol, one always ends up with pieces of material with different structural properties. Is it important to know whether these structural differences have any important effect on the physical observables of the glass, or, in other words, which observables would \emph{self average} such that their sample-to-sample fluctuations would be negligible in the thermodynamic limit. \emph{Self-averaging} assumptions go a long way back, at least to Tool's first work on fictive temperatures~\cite{Tool46}, and are a basic underlying assumption in the field of study of the thermodynamics of disordered systems in general~\cite{pedestrians}, beyond structural glasses. As a matter of fact, self-averaging can be shown to be rigorously realized (at least for systems with short-range interactions) for extensive quantities as a consequence of the Central Limit Theorem~\cite{pedestrians}. From an experimental point of view, this means that if one measured an extensive observable (say the internal energy, or the thermal capacity) in a given glass, the result would be representative of all the glasses manufactured with the same protocol. From a theoretical point of view, this means that some properties of glassy states can be safely computed by averaging them over the amorphous structures available~\cite{corrado2,RainoneUrbani15}.
The assumption of self-averaging is not sufficiently scrutinized for intensive variables. While some observables strictly related to the structure of the glass, such as the refractive index~\cite{Ritland56} do not self-average, it is still a common assumption that all thermodynamic quantities, whether extensive or intensive, should share this property.

In this Letter we show that this expectation is {\em not} met in the case of the non-linear elastic coefficients~\cite{LandauElasticity} of a model molecular glass at \emph{all temperatures} below the glass transition. This leads to a breakdown of the elastic theory for the material. It had been shown before that this is the case for amorphous solids at zero temperature~\cite{HentschelKarmakar11}, but one could think however that temperature fluctuations may destroy the relevance
 of the findings at $T=0$. We show in this Letter that it is not so: the presence of anomalous sample-to-sample fluctuations of non-linear elastic coefficients leads to a breakdown of elasticity theory also in amorphous solids at experimentally and practically relevant temperatures.

{\bf Expressions of elastic coefficients}:
Let us consider a standard elasticity theory for a solid under simple shear strain (with $\gamma_{xy} = \gamma$ the only non-zero component of the strain tensor). This is written in the form of a Taylor expansion around zero strain~\cite{LandauElasticity}:
\begin{equation}
 \sigma(\gamma) = B_1\gamma + \frac{1}{2!}B_2\gamma^2 + \frac{1}{3!}B_3\gamma^3 + \dots
\end{equation}
where $\sigma = \sigma_{xy}$ is the only non-zero component of the stress tensor and
\begin{equation}
 B_n \equiv \left.\frac{d^n\sigma}{d\gamma^n}\right|_{\gamma=0}.
\end{equation}
$B_1$ is the usual shear modulus that is usually denoted as $\mu$, $\mu\equiv B_1$.
In a thermal setting, the stress can be written as a canonical ensemble average~\cite{HessKroger97,Lutsko89,WittmerXu13}
\begin{equation}
 \sigma(\gamma) \equiv \frac{1}{V}\left<\frac{dU}{d\gamma}\right> = \frac{1}{V}\frac{1}{Z(\gamma)}\int_{X\in\alpha(R)}dX\ \frac{dU}{d\gamma}e^{-\beta U_\gamma(X)},
 \label{eq:sigma}
\end{equation}
where as usual $\beta \equiv \frac{1}{k_B T}$ and $V$ is the system's volume; $U(X)$ is the system's potential energy and the strain is implemented through an affine transformation of particle coordinates~\cite{HessKroger97}. The canonical average will be replaced below by a time average, using time intervals $\tau$ for which the variables measured reach a stationary value, but with $\tau$ being much shorter than the glass relaxation time
(denoted usually as $\tau_\alpha$). This time interval allows the system to visit a \emph{restricted} domain $\alpha(R)$ of configurations; accordingly the integral is computed over this set of configurations which are visited by the glass particles which are confined around an amorphous structure $R$~\cite{corrado2,DubeyProcaccia16}.

To compute the elastic coefficients, one needs only to take derivatives of Eq.~\eqref{eq:sigma} with respect to the strain. Notice how in Eq.~\eqref{eq:sigma} the strain parameter is contained in the derivative $\frac{dU}{d\gamma}$, in the Boltzmann factor, and in the partition function $Z(\gamma)$. When taking further derivatives of the derivative term, one will in general get a term of the kind $\left<\frac{\partial^n U}{\partial \gamma^n}\right>$, while derivatives of the partition function and Boltzmann factor will yield cumulants of the stress and additional covariance terms. The shear modulus for example has the expression~\cite{DubeyProcaccia16}:
\begin{equation}
\mu \equiv B_1= \frac{1}{V}\left<\frac{\partial ^2U}{\partial \gamma^2}\right> - \beta V[\left<\sigma^2\right> - \left<\sigma\right>^2],
\label{B1}
\end{equation}
which is the sum of a generalization of the \emph{Born term} found in crystalline solids~\cite{Born98} and thermal fluctuations of the stress. For the first non-linear coefficient $B_2$ one has
\begin{equation}
\begin{split}
 B_2 =\ &\frac{1}{V}\left<\frac{\partial^3U}{\partial\gamma^3}\right> -3\beta V[\left<\sigma'\sigma\right>-\left<\sigma'\right>\left<\sigma\right>]\\ &+ (\beta V)^2\left<(\sigma-\left<\sigma\right>)^3\right> \ , \label{B2}
 \end{split}
\end{equation}
where we have used the compact notation $\sigma' = \frac{\partial \sigma}{\partial \gamma}$. In the appendix we derive the expressions for the nonlinear coefficients up to 3rd order.
Since these coefficients are computed by sampling a glassy space of configurations selected by an amorphous structure, their values will depend on the particular glass sample under consideration, as detailed in the Introduction. We are interested in their probability distribution over samples, and in particular in sample-to-sample fluctuations
\begin{equation}
\overline{(\delta B_n)^2} \equiv \overline{(B_n-\overline{B_n})^2}\ , \label{var}
\end{equation}
where $\overline{(\bullet)}$ denotes the average over samples. Naive Central Limit Theorem considerations would suggest
$
\overline{(\delta B_n)^2} \simeq \frac{1}{V}\ ,
$
which would imply self-averaging. In the following we present evidence that this assumption fails for all $n\geq 2$.

{\bf Numerical simulations}:
We compute the elastic coefficients $B_n$ up to 3rd order from Molecular Dynamics (MD) simulations of a Kob-Andersen~\cite{94KA} 65/35 binary mixture in two dimensions. The Lennard-Jones potentials used are detailed in the SI. We always start by simulating the liquid at $T=0.4$, whereupon the relaxation of the binary correlation function is still exponential. Next we cool the
system at a rate of $10^{-6}$ in Lennard-Jones time units, as explained in the appendix, to the final target temperature of
$T=10^{-6}$. The system is now heated up instantaneously to a working temperature in the range $T\in [0.05,0.35]$ in
steps of $0.05$. The system is then ``equilibrated" by running 100,000 MD steps. To measure any desired quantity we
now run $\tau=200,000$ MD steps and measure the time average of the said quantity. Thus for example if we want to measure
$\langle \sigma^4 \rangle$ we compute
\begin{equation}
\langle \sigma^4 \rangle \equiv \tau^{-1}\sum^\tau_{i=1} \sigma^4(t_i) \ ,
\end{equation}
where $t_i$ are the MD steps.
\begin{figure}
\includegraphics[scale = 0.23]{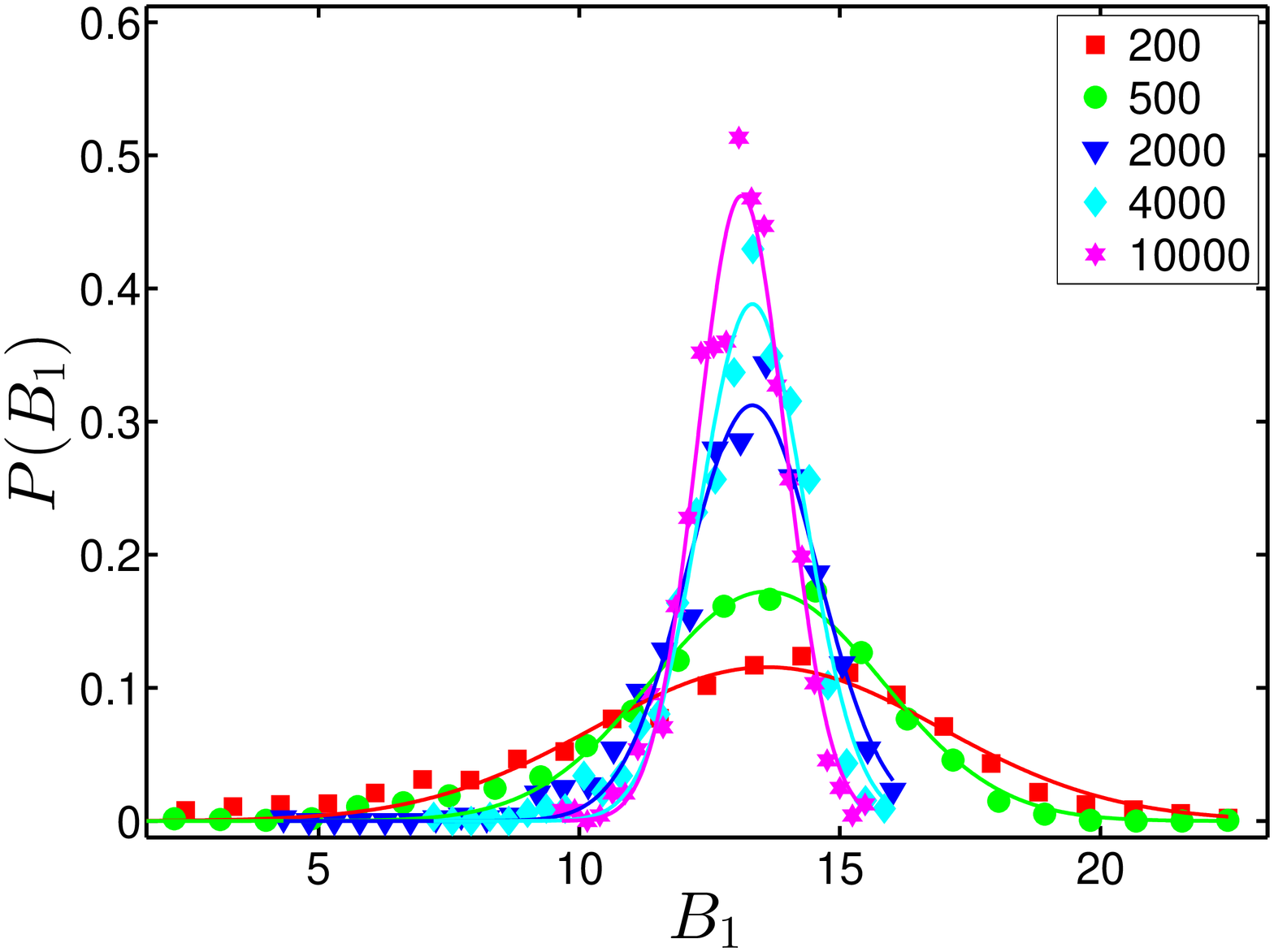}
\includegraphics[scale = 0.23]{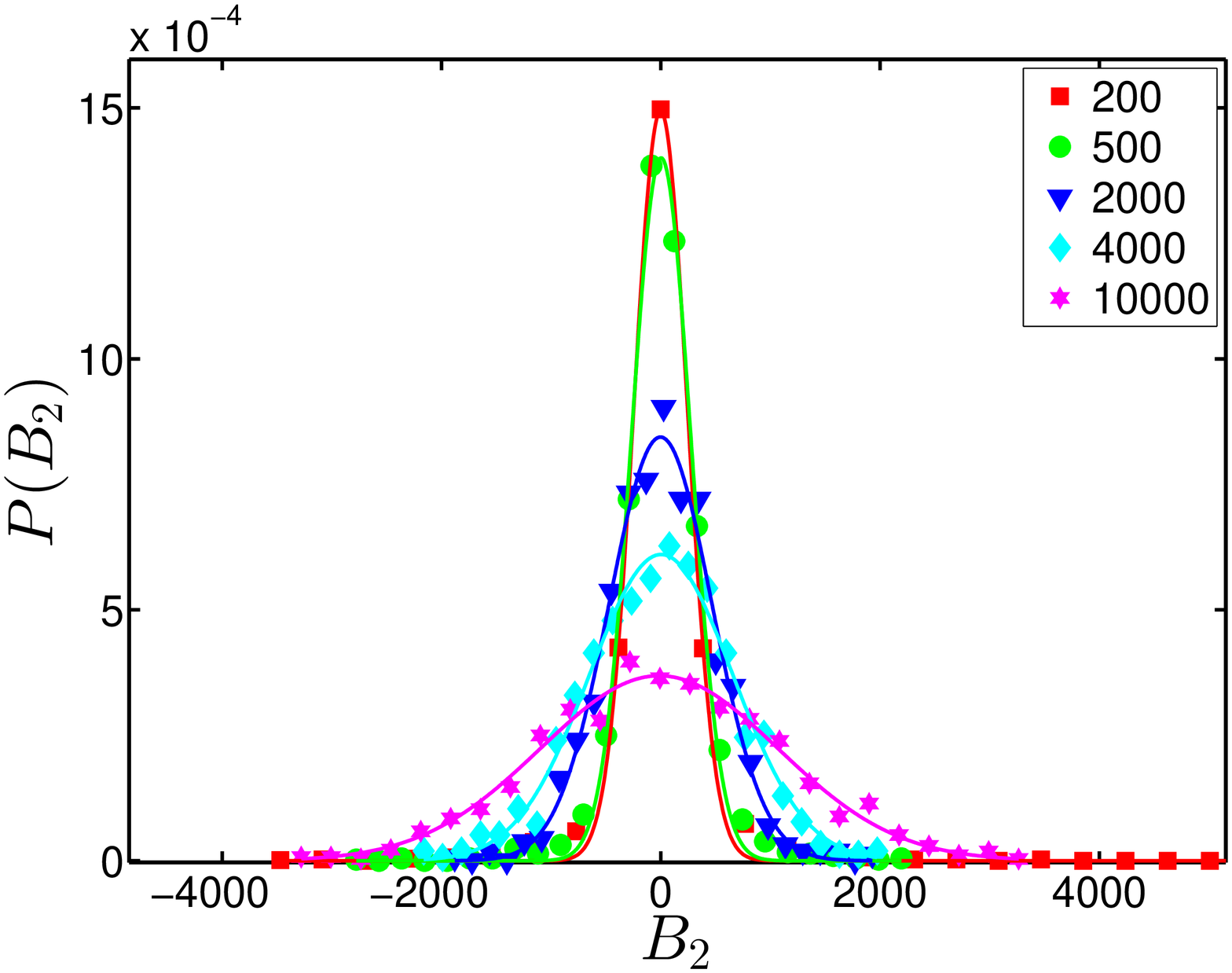}
\includegraphics[scale = 0.23]{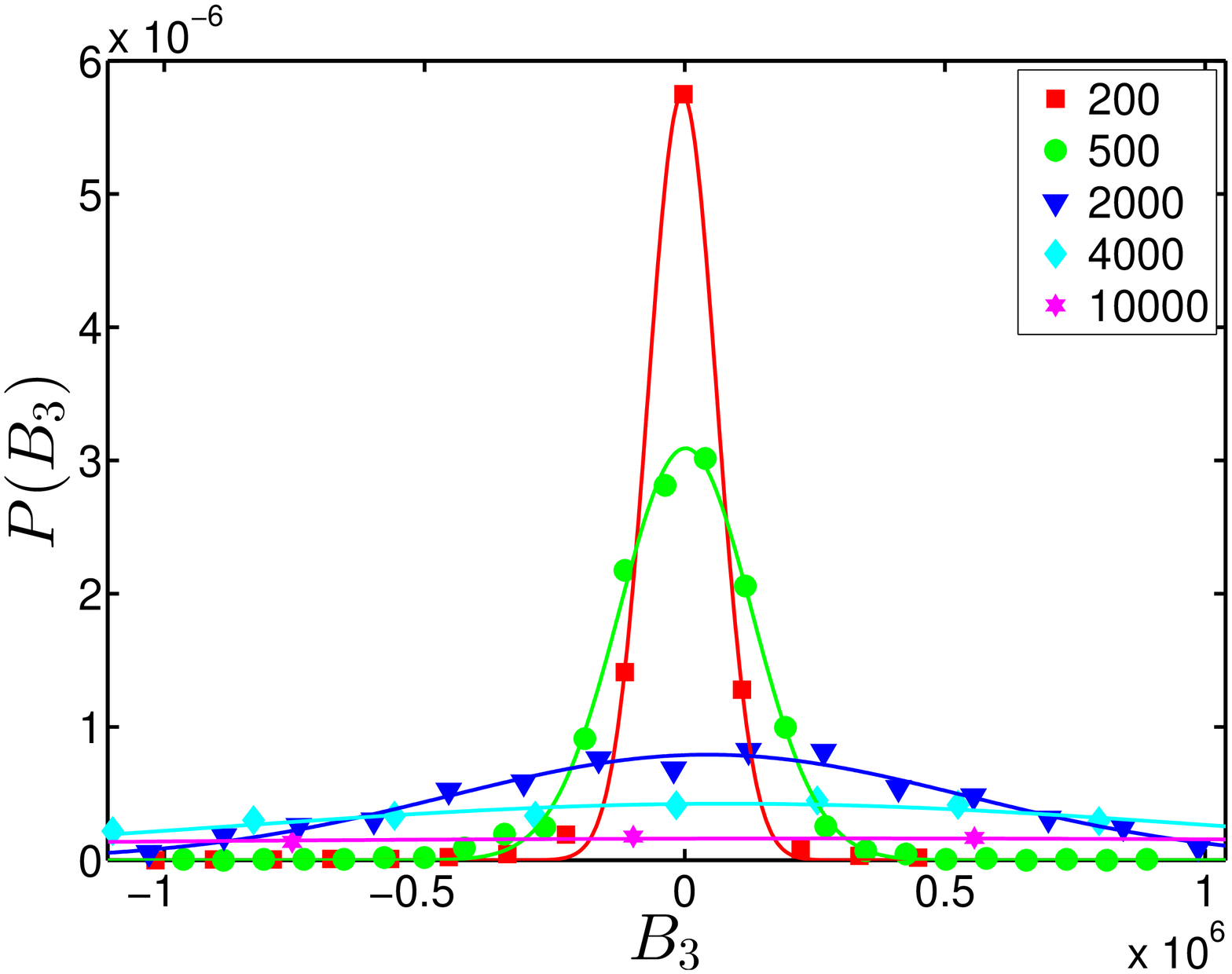}
\caption{The distributions of $B_1=\mu$, $B_2$ and $B_3$ over the realizations for $T=0.15$, for system
sizes from $N=200$ to $N=10000$. The lines are Gaussian fits to the data, from which we compute the variances. 
The distribution of the shear
modulus sharpens when the system size increases. The distributions of $B_2$ and $B_3$ broaden with increasing 
system size, refuting any hope for self-averaging. The distributions of higher order coefficients broaden faster and faster.}
\label{pdfs}
\end{figure}
Having computed the wanted quantity in this way, we repeat the process 1000 times, using different initial configurations
from the run at $T=0.4$, each of which will yield a different glass sample, or realization. The found values are histogrammed and normalized to yield a probability distribution function (pdf).
This pdf is now used to evaluate the average over the 1000 samples and the variance, Eq.~(\ref{var}). Our numerical setup is thus equivalent to the production of an ensemble of glass samples, each manufactured with the same, exactly reproduced protocol.

\begin{figure}
\includegraphics[scale = 0.20]{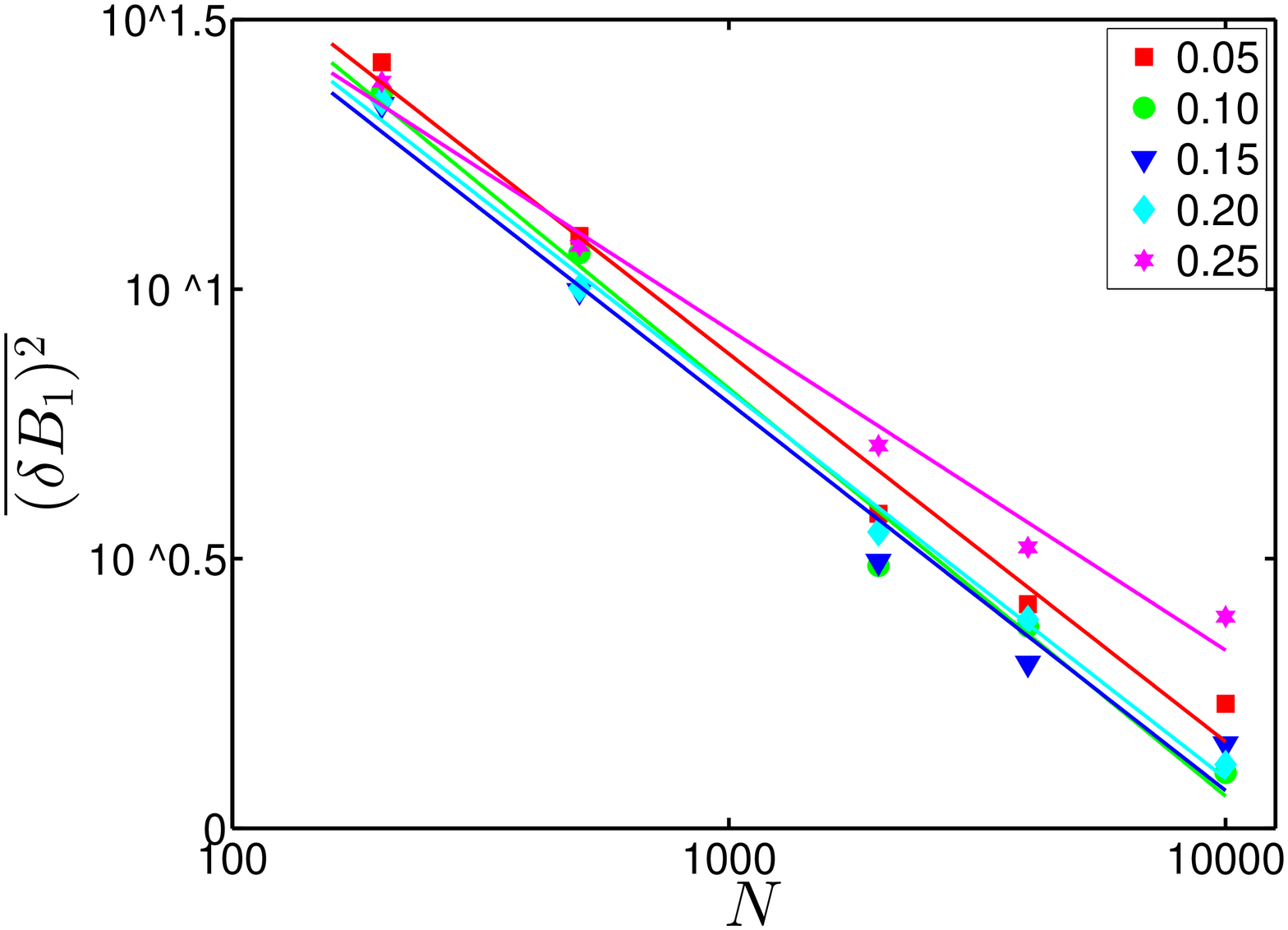}
\includegraphics[scale = 0.20]{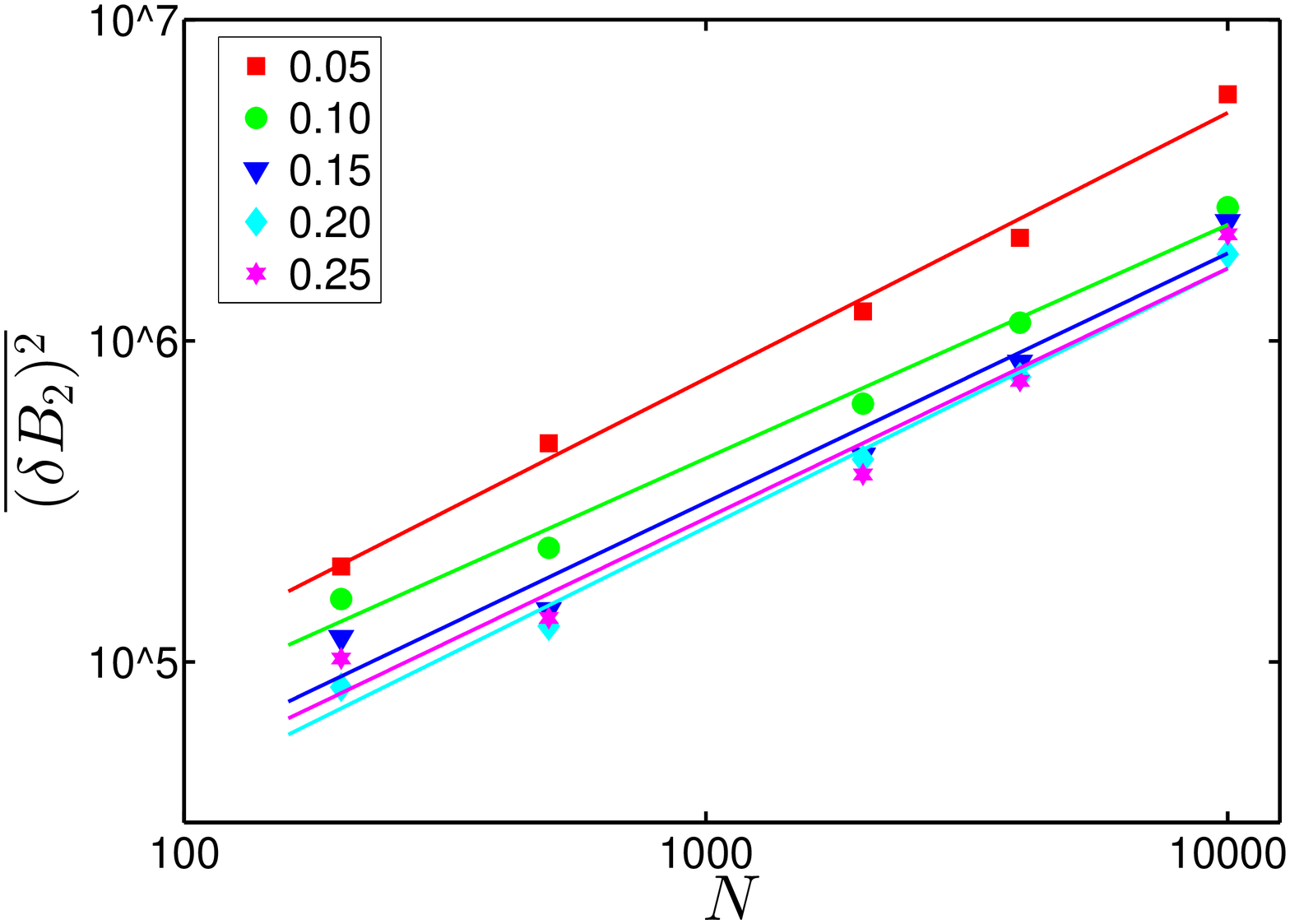}
\includegraphics[scale = 0.20]{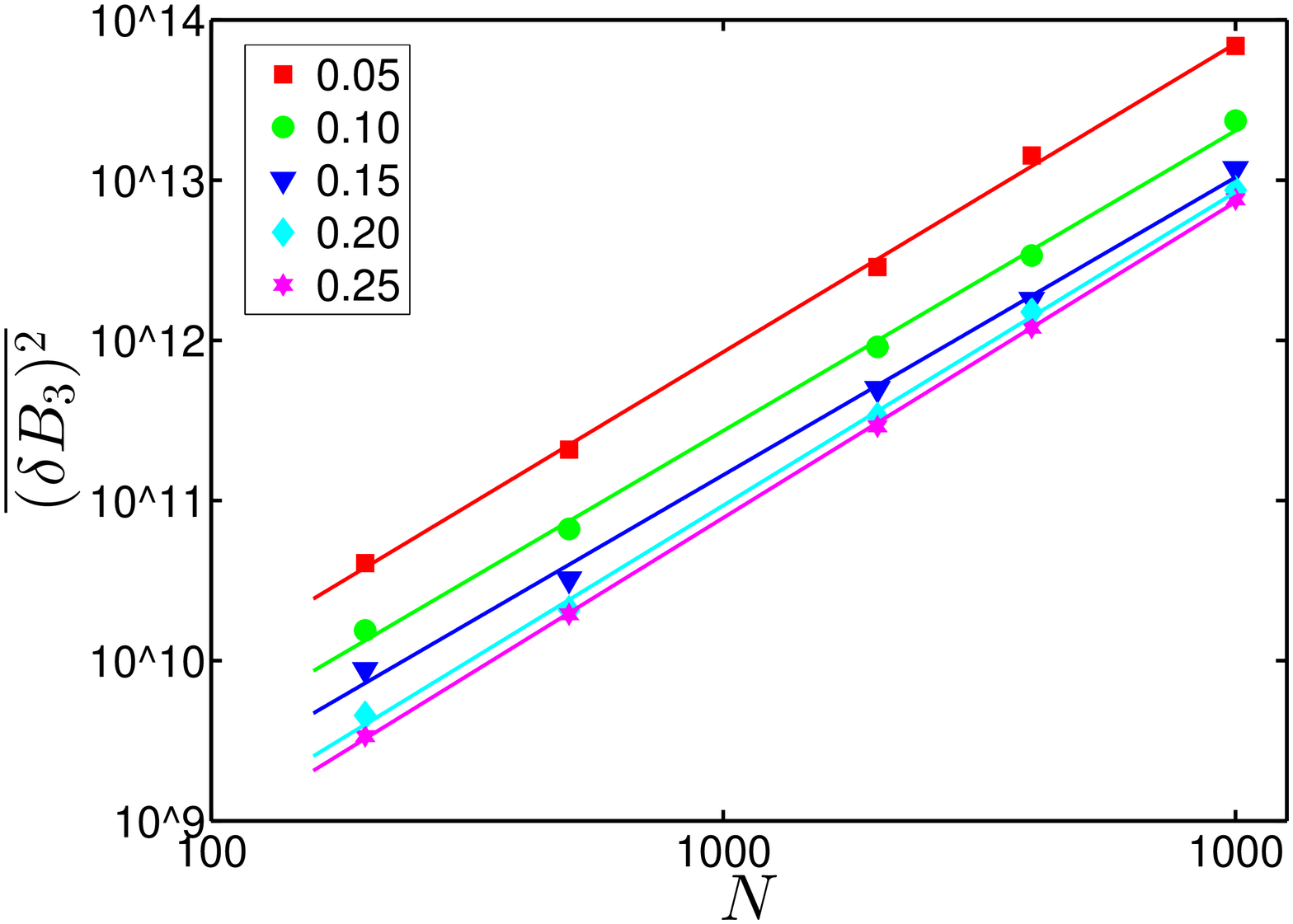}
\caption{The variances of $B_1, B_2$ and $B_3$ over the realizations as a function of system
size for different temperatures. The lines are least-square fits to the data. Note to within
the available accuracy the system-size dependence of the variances appears temperature independent, at least
up to $T=0.25$.}
\label{variance}
\end{figure}

{\bf Results}: a representative set of results for the distributions of $B_1, B_2$ and $B_3$ over the realizations is shown
in Fig.~\ref{pdfs} for $T=0.15$. Similar results are seen for all the temperature range: the distribution of the shear
modulus over the realizations sharpens with the system size, indicating self averaging in the thermodynamic
limit. The distributions of $B_2$ and $B_3$ (and in fact of all $B_n$ with $n\ge 2$) broaden
rapidly with increasing system sizes, indicating a breakdown of self-averaging and of nonlinear elasticity. The rate of broadening of the distributions increases with the order of the coefficient under consideration. As an example, let us consider the variances of the distributions of the first three moduli; to evaluate their finite-size scaling, we perform Gaussian least-squares fits of the data and consider the resulting variances, which are shown in Fig~\ref{variance} as a function of the system size at different temperatures.
Denoting the variance of $B_k$ as $\overline{(\delta B_k)^2}$ we find that
\begin{equation}
\overline{(\delta B_1)^2} \sim N^{\alpha_1} \ ,  \overline{(\delta B_2)^2} \sim N^{\alpha_2} \ , \overline{(\delta B_3)^2} \sim N^{\alpha_3} \ ,
\end{equation}
With $\alpha_1=-0.68\pm0.08$,  $\alpha_2=0.78\pm 0.05$ and $\alpha_3= 1.92\pm 0.06$ independently of the temperature in the range $T\in[0.05,0.25]$. 

\begin{figure}
\includegraphics[scale = 0.45]{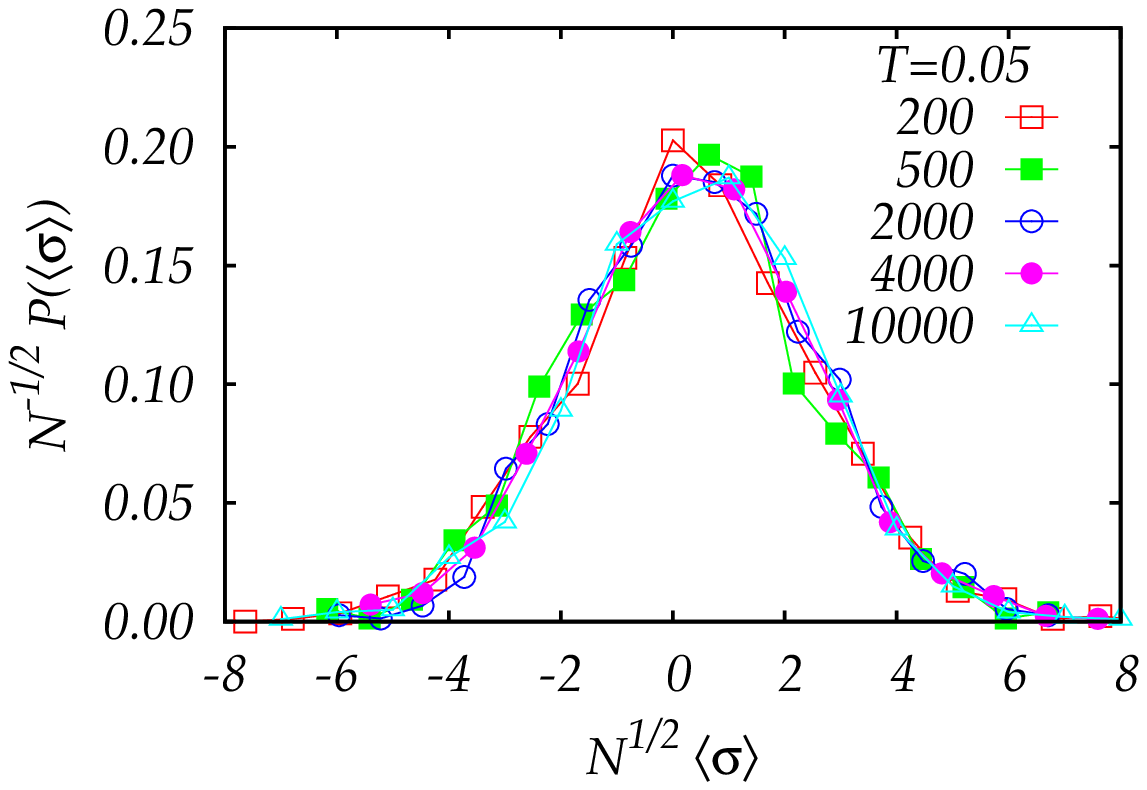}
\includegraphics[scale = 0.45]{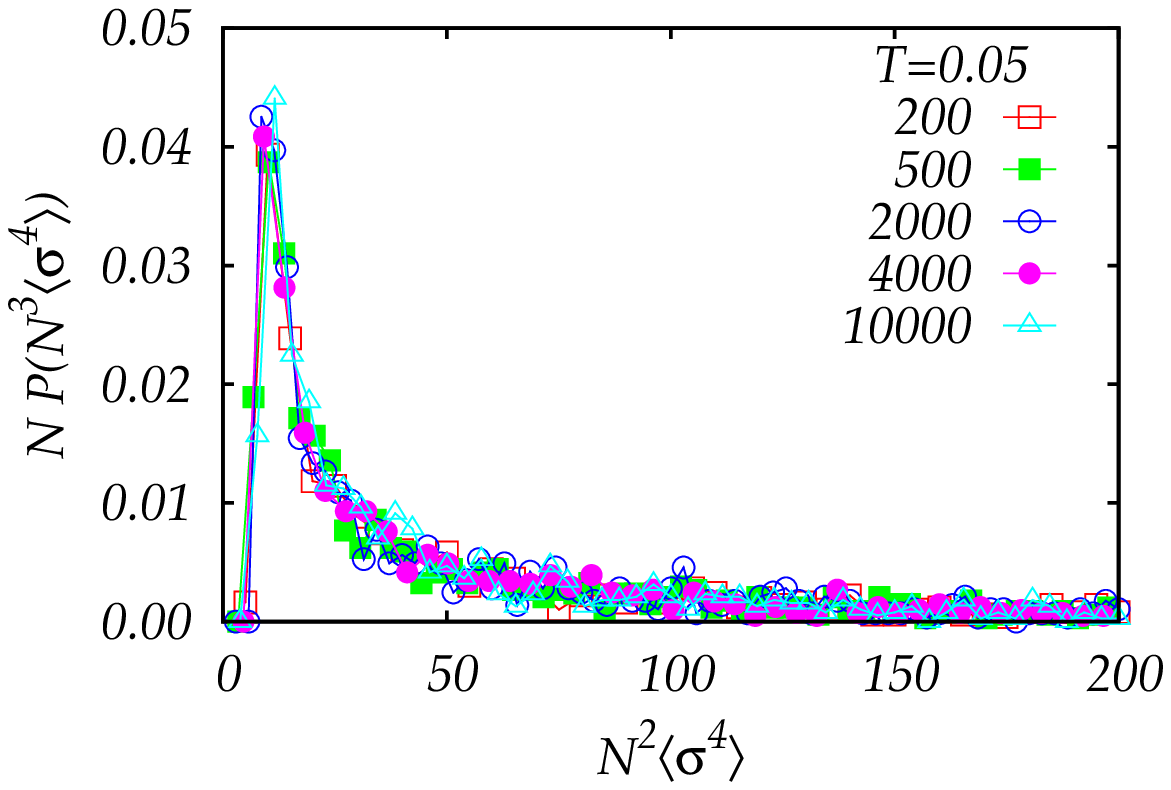}
\includegraphics[scale = 0.45]{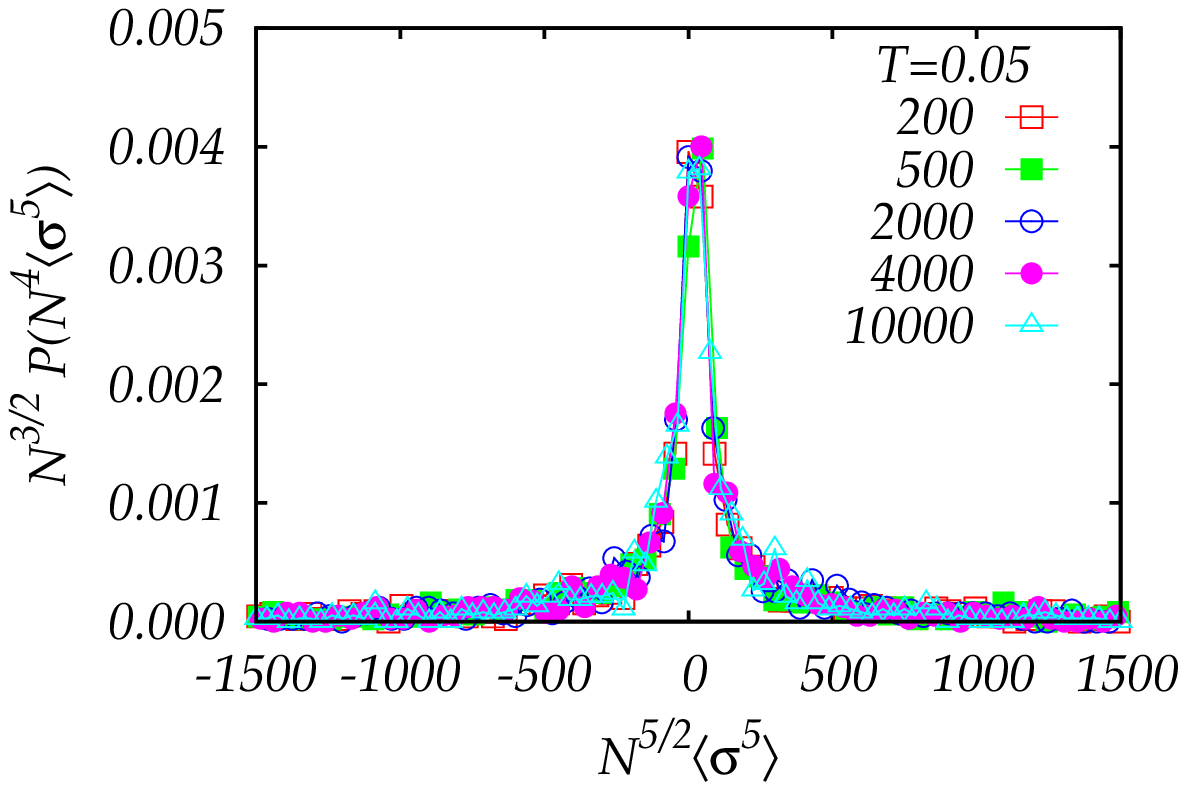}
\caption{Examples of rescaled pdf's of $X_k$ for $k=1,4$ and 5, multiplied by $N^{k/2-1}$ and plotted
as a function of $N^{k/2} \langle \sigma^k\rangle$ at $T=0.05$. Similar data collapses are found for
other $k$ values and for other temperatures without changing the exponents. Similar data collapse for $k=2,3$ and 6
can be found in the appendix.}
\label{pdfmom}
\end{figure}

To shed light on the breakdown of self-averaging it is useful to consider the sample-to-sample fluctuations
of the moments of the stress. We note that the fluctuations in the Born-like terms in any of the $B_k$ moduli are
always convergent. The reason for divergence are the moments $\langle \sigma^k\rangle$ which appear in the expressions
for the coefficients $B_k$, multiplied by a suitable factor of $V^{k-1} \sim N^{k-1}$ to make all the $B_k$s intensive. It is therefore interesting to consider the
sample-to-sample fluctuations of $X_k\equiv N^{k-1}\langle\sigma^k\rangle$. Accordingly we consider the pdf's of $P(X_k)$ over
our glass samples.
In Fig.~\ref{pdfmom} we show representative results of these pdf's in a rescaled form. The upshot
of the analysis is that we can collapse the data of these pdf's for different system sizes if we plot
$N^{k/2-1}P(X_k)$ as a function of $N^{k/2} \langle \sigma^k\rangle$.
It is an immediate exercise to evaluate therefore the system size dependence of the variance of $X_k$, denoted here as $\overline{(\delta X_k)^2}$, obtaining the scaling dependence
\begin{equation}
\overline{(\delta X_k)^2} \sim N^{k-2} \ ,
\end{equation}
independently of the temperature. It now becomes clear that the convergent result for $\overline{(\delta B_1)^2}$ and the fact that $\alpha_2<1$ and $\alpha_3<2$ must follow from a cancellation of the leading $N$ dependence in the terms involving the stress fluctuations.
A precise determination of the scaling exponents and the question of their universality or non-universality must await
a very extensive set of numerical simulations which are outside the scope of this Letter.

{\bf Discussion}: It is interesting to examine the correspondence between the divergence of the variances of the nonlinear elastic
coefficients at $T=0$ and at finite temperature. At $T=0$ the expression for the shear modulus, instead of Eq.~\ref{B1}, reads
\cite{99ML}
\begin{equation}
B_1(T=0)=\frac{1}{V}\frac{\partial^2 U}{\partial \gamma^2}-  \frac{1}{V} \boldsymbol{\Xi} \cdot \boldsymbol{\mathcal{H}}^{-1}\cdot \boldsymbol{\Xi}  \ ,
\label{B1T0}
\end{equation}
where $\boldsymbol{\mathcal{H}}$ is the Hessian matrix and $\boldsymbol{\Xi} \equiv \partial^2 U/\partial \boldsymbol{r}_i\partial \gamma$.
Higher order nonlinear moduli contain three, five, and more factors of $\boldsymbol{\mathcal{H}}^{-1}$ and their sample to sample
fluctuations stem from the existence of arbitrarily small eigenvalues of the Hessian matrix when the system
size increases\cite{HentschelKarmakar11}. It can be proven that the stress fluctuation term in Eq.~\ref{B1} approaches smoothly the second term in Eq.~(\ref{B1T0}), and that the cumulant terms in $B_k(T)$ approach in the same way the analogous
term of the athermal counterpart, see \cite{10KLP} and appendix. In recent years, much research has been devoted to the concept of \emph{marginality}~\cite{15MW} in disordered systems,
which can be broadly defined as the possibility to destabilize a system with a generic perturbation without having to pay an energy cost. In the case of athermal systems, such as jammed packings,
those perturbations are \emph{mechanical} in nature (for example, the opening of a contact between two grains in a packing) and marginality manifests under the guise of arbitrarily low-lying eigenvalues in the Hessian of the system, related to floppy modes that can be excited with no energy cost.
As detailed in \cite{HentschelKarmakar11}, they are precisely those modes that cause the breakdown of non-linear elasticity in athermal glasses. The correspondence between the second term in Eq~(\ref{B1T0}) and the stress fluctuations of the thermal case (which, as we pointed out, cause the breakdown of non-linear elasticity in the present case) highlights how the \emph{mechanical} marginality found in athermal amorphous systems must have a thermal, thermodynamic counterpart in terms of the presence of anomalous thermal fluctuations, which in turn induce, through the Fluctuation-Dissipation theorem, an anomalous response of the system to even arbitrarily small thermodynamic perturbations, such as strain or a magnetic field. We argue that a better understanding of the links between mechanical and thermal marginality is paramount for the final achievement of a complete and consistent theoretical picture of the physics of amorphous systems.

Finally we should relate these findings to a recent theoretical work~\cite{16BU} predicting a so-called Gardner transition~\cite{Ga85} in
thermal glass forming liquids \cite{fullRSB,16BU}. Fundamentally the prediction is that at some temperature, lower
than the glass transition temperature, there should be a qualitative change in the nature of the free-energy
landscape, generating a rough scenery with arbitrarily small barriers between local minima. The connection to the
present work is that this phenomenon is accompanied by a breakdown of nonlinear elasticity in much the same
way reported above. The available theory pertains to a mean field treatment and comparison of exponents is
probably not warranted. Nevertheless it is interesting that the shear modulus is expected to exist, and the
variances of $B_k$ with $k\ge 3$ are expected to diverge with the system size, in agreement with the predictions
of Ref.~\cite{HentschelKarmakar11} and the findings of the present Letter.  In Ref.~\cite{16BU} it is also predicted
that the phenomenon should disappear when the system is heated above the (protocol dependent) Gardner temperature,
a claim that we are not in position to confirm or refute. A careful search of a putative Gardner temperature
would require repeating our analysis on extremely slowly quenched glasses as a way to provide a good separation
of the Gardner point and the point of disappearance of the shear modulus~\cite{16BU}. Such an analysis is beyond the scope
of the present Letter but appears to be a worthwhile endeavor for future research.

\acknowledgments
This work had been supported in part by an ERC ``ideas" grant STANPAS and by the Minerva Foundation,
Munich Germany. We benefited from useful discussions with Giulio Biroli and Pierfrancesco Urbani.

\clearpage

\appendix

\begin{widetext}

\section{Expressions of the elastic coefficients}

We present here the expressions of the elastic coefficients that are studied in the main text. We start from the definition of the stress
\begin{equation}
\sigma =
\frac{1}{V} \left[ \frac{1}{Z(\gamma)} \int \frac{dU}{d\gamma} e^{-\beta U(\gamma)} dX \right],
\end{equation}
where
\begin{equation}
Z(\gamma )= \int e^{-\beta U(\gamma)} dX \quad.
\end{equation}
We now take a derivative of this expression with respect to $\gamma$, which will be by definition equal (once computed at $\gamma=0$) to the shear modulus. We get
\begin{align*}
\frac{d \sigma}{d\gamma}  &=
\frac{1}{V} \left[  \frac{1}{Z(\gamma)} \int\frac{\partial^2 U}{\partial\gamma^2} e ^{-\beta U(\gamma)} dX
-\beta \frac{1}{Z(\gamma)} \int \left( \frac{\partial U}{\partial \gamma} \right)^2 e ^{-\beta U(\gamma)} dX
+\beta \frac{1}{Z(\gamma)^2} \left(  \int  \frac{\partial U}{\partial \gamma}  e ^{-\beta U(\gamma)} dX \right)^2  \right],
\end{align*}
since
\begin{equation}
\frac{\partial}{\partial \gamma}Z(\gamma )= -\beta \int\frac{\partial U}{\partial \gamma} e^{-\beta U(\gamma)} dX;
\end{equation}
now, since $\sigma$ is an intensive quantity, $\sigma \equiv \frac{1}{V}\thav{\frac{\partial U}{\partial\gamma}}$, we have to multiply the last two terms by $\frac{V}{V}$, and we finally get
\begin{equation}
\mu = \frac{1}{V}\thav{\frac{\partial^2 U}{\partial\gamma^2}} - \beta V[\thav{\sigma^2}-\thav{\sigma}^2],
\end{equation}
as reported in the main text and in~\cite{DubeyProcaccia16}. We now take further derivatives in order to compute the 2nd- and 3rd-order coefficient. For the second derivative we have
\begin {align*}
\frac{d^2 \sigma}{d \gamma^2}   =\ &
\frac{1}{V} \Bigg[
\frac{1}{Z} \int\frac{\partial^3 U}{\partial^3 \gamma} e^{-\beta U} dX \\&
+\beta \frac{1}{Z^2} \int  \frac{\partial^2 U}{\partial \gamma^2}  e^{-\beta U} dX
 \int  \frac{\partial U}{\partial \gamma}  e^{-\beta U} dX
-3\beta \frac{1}{Z}  \int  \frac{\partial U}{\partial \gamma} \frac{\partial^2 U}{\partial \gamma^2}  e^{-\beta U} dX \\&
-\beta^2  \frac{1}{Z^2}   \int  \frac{\partial U}{\partial \gamma} e^{-\beta U} dX  \int  \frac{\partial^2 U}{\partial \gamma^2}  e^{-\beta U} dX \\&
+\beta^2  \frac{1}{Z} \int \left( \frac{\partial U}{\partial \gamma} \right)^3 e^{-\beta U} dX
\\&
+2\beta  \left(  \frac{1}{Z} \int  \frac{\partial U}{\partial \gamma}  e^{-\beta U} dX \right) 		\bigg(
 		\frac{1}{Z} \int  \frac{\partial^2 U}{\partial \gamma^2}  e^{-\beta U} dX
 		- \frac{\beta}{Z(\gamma)} \int \left( \frac{\partial U}{\partial \gamma} \right)^2 e^{-\beta U} dX
 		+\frac{\beta^2 }{Z(\gamma)^2} \left(  \int  \frac{\partial U}{\partial \gamma}  e^{-\beta U} dX \right)^2
  	\bigg)
\Bigg];
\end{align*}
and once taken care of the volume factors, we get the final result for $B_2$
\begin{equation}
\begin{split}
 B_2 =\ &\frac{1}{V}\left<\frac{\partial^3 U}{\partial\gamma^3}\right> -3\beta V[\left<\sigma'\sigma\right>-\left<\sigma'\right>\left<\sigma\right>] + (\beta V)^2\left<(\sigma-\left<\sigma\right>)^3\right> \ , \label{B2}\\
 =\ &\frac{1}{V}\left<\frac{\partial^3 U}{\partial\gamma^3}\right> -3\beta V \Cov[\sigma',\sigma] + (\beta V)^2 \kappa_3[\sigma]
 \end{split}
\end{equation}
as reported in the main text. Higher order coefficients can be computed with the same method, and, even though the expressions become longer and cumbersome, the calculation in itself is trivial. The result for $B_3$ for example is:
\begin{align*}
 B_3=\ &\frac{1}{V}\left< \frac{\partial^4 U}{\partial \gamma^4}\right>  \\&
		+3 \beta V \langle \sigma'\rangle ^2
		-3 \beta V  \langle (\sigma')^2\rangle \\&
		+4\beta V  \langle \sigma''\rangle \langle  \sigma\rangle
		-4\beta V  \langle \sigma'' \sigma\rangle \\&
		+6\beta^2 V ^2 \langle \sigma^2 \sigma' \rangle
		-6\beta^2 V ^2 \langle \sigma^2\rangle \langle \sigma' \rangle \\&	
		+12\beta^2 V ^2 \langle \sigma\rangle^2 \langle  \sigma' \rangle
		-12\beta^2 V ^2 \langle \sigma\rangle \langle  \sigma \sigma' \rangle  \\&
		+ \beta^3 V ^3 ( 4\langle \sigma^3\rangle \langle \sigma\rangle
		+3 \langle \sigma^2\rangle ^2 -12 \langle \sigma^2\rangle  \langle \sigma\rangle ^2
		+6 \langle \sigma\rangle ^4 -\langle \sigma^4\rangle)  \\
		=\ &\frac{1}{V}\left< \frac{\partial^4 U}{\partial\gamma^4}\right> -3\beta V[\thav{(\sigma')^2}-\thav{\sigma'}^2] -4V\beta [\thav{\sigma''\sigma}-\thav{\sigma''}\thav{\sigma}]
	+6V^2 \beta^2 [\thav{\sigma'\sigma^2}-\thav{\sigma'}\thav{\sigma^2}]
	\\&  -12 V^2\beta^2 \thav{\sigma}[\thav{\sigma\sigma'}-\thav{\sigma}\thav{\sigma'}]
	+3V^3\beta^3 (\thav{\sigma^2}-\thav{\sigma}^2)^2
	- V^3\beta^3 \thav{(\sigma-\thav{\sigma})^4}.\\
	=\ &\frac{1}{V}\left< \frac{\partial^4 U}{\partial\gamma^4}\right> -3 V\beta \Var[\sigma'] -4V\beta \Cov[\sigma'',\sigma] + 6V^2\beta^2\Cov[\sigma',\sigma^2]\\
	&-12 V^2\beta^2 \Exp[\sigma]\Cov[\sigma,\sigma'] +3V^3\beta^3(\Var[\sigma])^2 - V^3\beta^3 \kappa^4[\sigma]
\end{align*}

\section{Details on the numerics}

\subsection{Model Details:}
We study the two-dimensional Kob-Andersen binary mixture with a 65:35 ratio of particles A and B,
where particles are point particles and interact via a shifted and smoothed Lennard-Jones (LJ) potentials, $u_{\alpha\beta}(r)$, given by
\begin{equation}
u_{\alpha\beta}(r) = \begin{cases} u^{LJ}_{\alpha\beta}+A_{\alpha\beta} +B_{\alpha\beta}r+C_{\alpha\beta}r^2, & \mbox{if } r \leq R^{cut}_{\alpha\beta}
                                 \\ 0, & \mbox{if } r > R^{cut}_{\alpha\beta}, \end{cases}
\label{Usmooth}
\end{equation}
where
\begin{equation}
u^{LJ}_{\alpha\beta} = 4\epsilon_{\alpha\beta}\left[\left(\frac{\sigma_{\alpha\beta}}{r}\right)^{12} - \left(\frac{\sigma_{\alpha\beta}}{r}\right)^6\right].
\label{ULJ}
\end{equation}
The smoothing of the potentials in Eq. (\ref{Usmooth}) is such that they vanish with
two zero derivatives at distances $R^{cut}_{\alpha\beta} = 2.5\sigma_{\alpha\beta}$.
The parameters for smoothing the LJ potentials in Eq. (\ref{Usmooth}) and
for A and B particle type interactions in Eq.(\ref{ULJ})\cite{94KA} are given in the following table
\begin{center}
\begin{tabular}{ |c|c|c|c|c|c| }
\hline
Interaction & $\sigma_{\alpha\beta}$ & $\epsilon_{\alpha\beta}$ & $A_{\alpha\beta}$ & $B_{\alpha\beta}$ & $C_{\alpha\beta}$ \\
 \hline
 AA & 1.00 & 1.0 & 0.4527 & -0.3100 & 0.0542 \\
 BB & 0.88 & 0.5 & 0.2263 & -0.1762 & 0.0350 \\
 AB & 0.80 & 1.5 & 0.6790 & -0.5814 & 0.1271 \\
 \hline
\end{tabular}
\end{center}
The reduced units for mass, length, energy and time have been taken as $m$,
$\sigma_{AA}$, $\epsilon_{AA}$ and $\sigma_{AA}\sqrt{m/\epsilon_{AA}}$ respectively.

\subsection{Simulation Details:}
All the simulations were carried out with Molecular Dynamics (MD) in NVT conditions, using a velocity-Verlet algorithm with a time step of
$\Delta t=$0.005 in reduced units. A Berendsen  thermostat, with a time constant of $5$ in reduced units,
was used to maintain the desired temperature. All simulations have been performed at constant density $\rho=1.162$
with system sizes ranging from $N$=200 to $N=$10000 and a temperature range from $T=$0.05 to $T=$0.35 with a gap of 0.05.

\subsection{Protocol for the Preparation of Amorphous Solids:}
 In order to prepare amorphous solids, we always start with a random configuration generated
 at $\rho=$1.162 and then equilibrate it at a high temperature $T=0.4$ for 400,000 MD steps. At this
 temperature correlation functions still decay exponentially and the system behaves like a liquid. Next, we cool down the system, with a cooling rate of
 $\Delta T=10^{-6}$ in reduced units, to a target temperature of $T=0.000001$. We repeat this process starting from
 different initial conditions at $T=0.4$ to generate the ensemble of
 1000 amorphous solids at each system size.

 \subsection{Data collapse for higher order moments}
To complement the data presented in Fig.~3 of the main text we report in Fig.~\ref{fig:collapse} the data collapse obtained with the scaling ansatz $N^{k/2-1}P(X_k) = f(N^{k/2}\thav{\sigma^k})$ for $k=2,3,6$.
 \begin{figure}
  \includegraphics[width=0.33\textwidth]{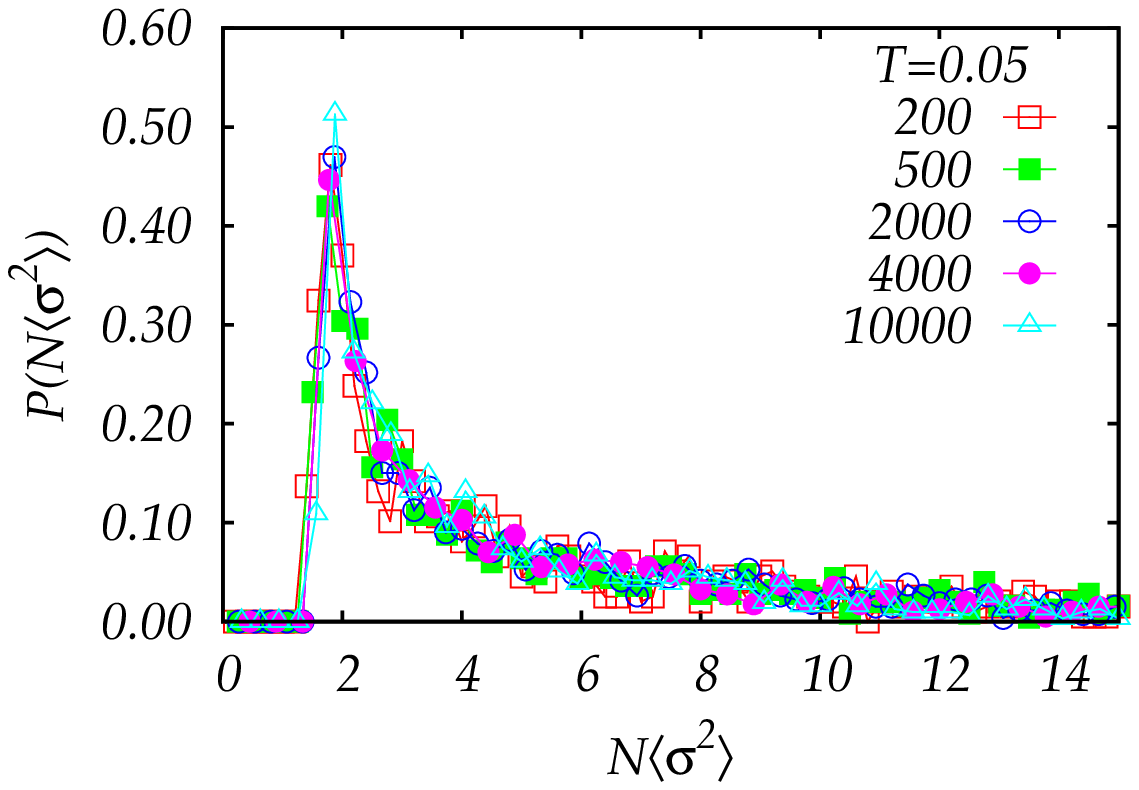}
  \includegraphics[width=0.33\textwidth]{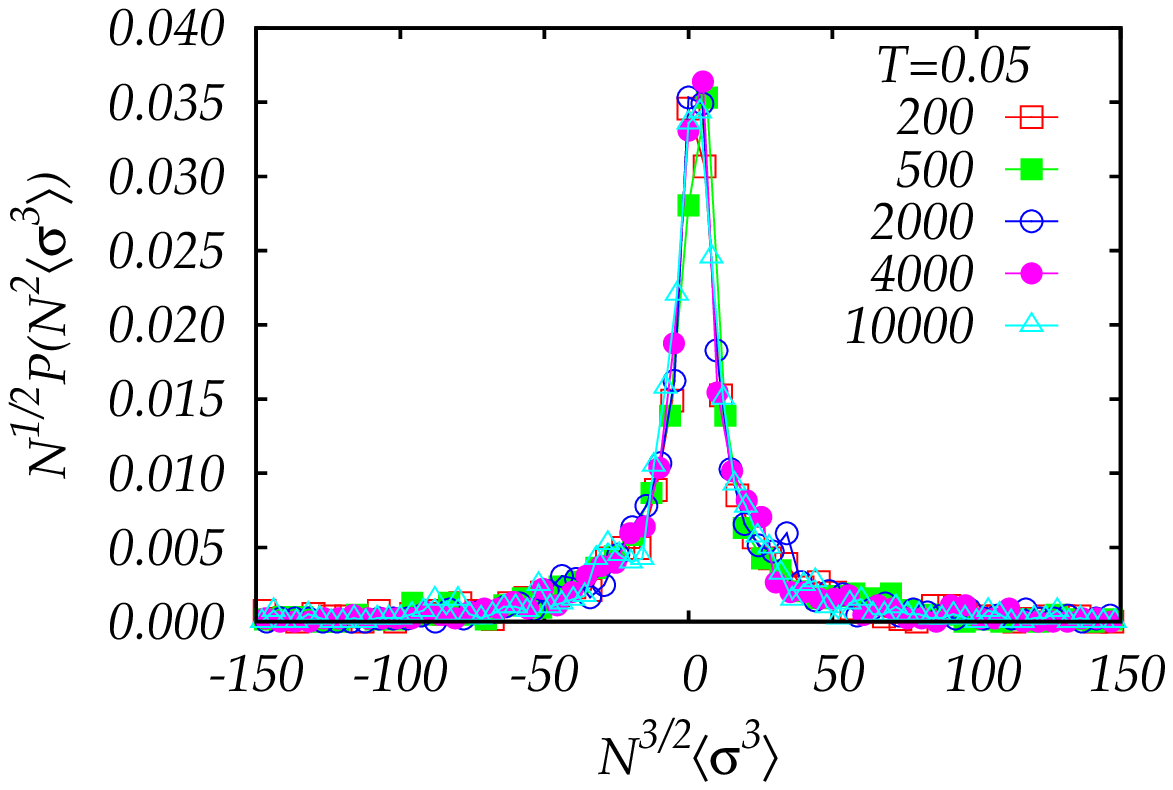}
  \includegraphics[width=0.33\textwidth]{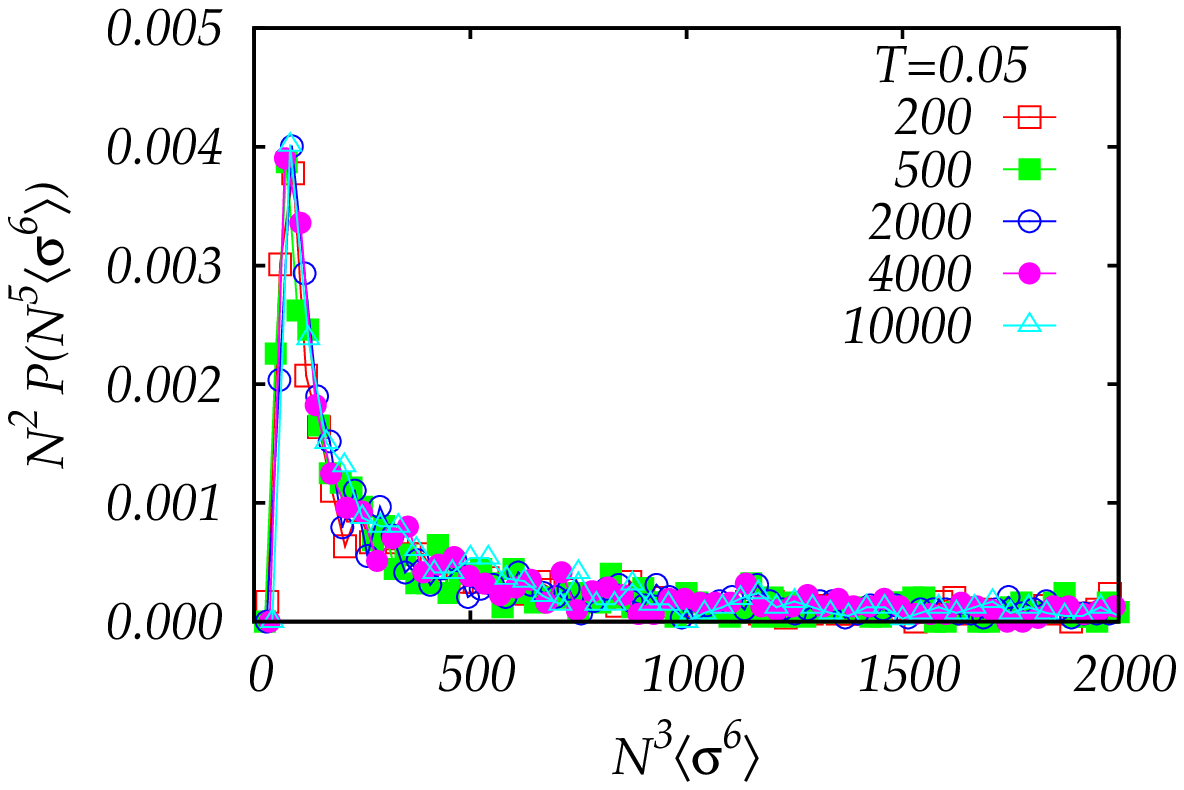}
  \caption{Data collapse of the $P(X_k)$ obtained with the scaling ansatz reported in the main text. Even though it is a purely phenomenological ansatz without a theoretical justification, the results are satisfying. \label{fig:collapse}}
 \end{figure}

\section{Low-temperature limit of thermal fluctuations}
We show here that, for two generic observables $A(X)$ and $B(X)$, one has
\begin{equation}
 \lim_{\beta\to\infty} \beta[\langle A(X)B(X)\rangle - \langle A(X)\rangle\langle B(X)\rangle] = \left.[\nabla A \cdot \H^{-1} \cdot \nabla B]\right|_{X=X^*}
\end{equation}
Where $\H^{-1}$ is the inverse Hessian of the system and $X^*$ is the inherent structure the systems settles in when $T\to 0$. The proof is already provided in~\cite{KarmakarLerner10} in the case of elastic coefficients, here we report a simpler derivation for two generic observables.
We start by considering the average
\begin{equation}
 \thav{A} = \frac{\int dX\ A(X) e^{-\beta U(X)}}{\int dX\ e^{-\beta U(X)}} =  \frac{\int dX\ e^{-\beta [U(X)-\frac{1}{\beta}\log A(x)]}}{\int dX\ e^{-\beta U(X)}}
\end{equation}
We compute the integrals with the saddle point method~\cite{AdvancedMethods}. Let us expand the arguments of the exponentials around the inherent structure. We get for the numerator:
\begin{equation}
\int dX\ A(X^*) \exp\{-\beta [U(X^*)-\frac{1}{\beta}\frac{1}{B}\nabla B\cdot\delta X+\frac{1}{2}\delta X\cdot\A\cdot \delta X + O(X^3)]\}
\end{equation}
and for the denominator
\begin{equation}
 \int dX\ \exp\{-\beta [U(X^*)+\frac{1}{2}\delta X\cdot\H\cdot\delta X + O(X^3)]\},
\end{equation}
where $\A$ is a matrix defined as
\begin{equation}
 \A \to \A_{i\alpha j\beta} \equiv \H_{i\alpha j\beta} -\frac{1}{\beta}\frac{\partial^2 \log A}{\partial x_{i\alpha}\partial x_{j\beta}}.
\end{equation}
where the Latin indexes denote particle coordinates and Greek ones spatial axes. The integral in the numerator in a Gaussian integral with a linear term, which can be straightforwardly computed. One gets
\begin{equation}
A(X^*)\exp\left[\frac{1}{2\beta}\left(\frac{\nabla A}{A} \cdot \A^{-1} \cdot \frac{\nabla A}{A}\right)\right]\sqrt{\frac{\pi}{\beta}}^{dN}\frac{1}{\sqrt{\det\A}},
\end{equation}
while the result for the denominator is
\begin{equation}
 \sqrt{\frac{\pi}{\beta}}^{dN}\frac{1}{\sqrt{\det\H}}
\end{equation}
where $d$ is the number of dimensions ($d=2$ in the present case, but the derivation is valid for any $d$);
in summary, we get for $\thav{A}$
\begin{equation}
 \thav{A} \simeq A(X^*)\exp\left[\frac{1}{2\beta}\left(\frac{\nabla A}{A} \cdot \A^{-1} \cdot \frac{\nabla A}{A}\right)\right] \sqrt{\frac{\det\H}{\det\A}},
\end{equation}
so in the $T\to 0$ limit we get, as expected
\begin{equation}
\lim_{T\to 0} \thav{A} = A(X^*).
\end{equation}

Let us now consider $\thav{AB}$ and $\thav{A}\thav{B}$. We get, using the same reasoning,
\begin{equation}
 \beta\thav{AB} \simeq \beta A(X^*)B(X^*)\exp\left[\frac{1}{2\beta}\left(\frac{\nabla A}{A}+\frac{\nabla B}{B}\right) \cdot \C^{-1} \cdot \left(\frac{\nabla A}{A} + \frac{\nabla B}{B}\right)\right] \sqrt{\frac{\det\H}{\det\C}},
\end{equation}
with the definition
\begin{equation}
\C \to \C_{i\alpha j\beta} \equiv \H_{i\alpha j\beta} -\frac{1}{\beta}\frac{\partial^2 \log A}{\partial x_{i\alpha}\partial x_{j\beta}} -\frac{1}{\beta}\frac{\partial^2 \log B}{\partial x_{i\alpha}\partial x_{j\beta}},
\end{equation}
while for the other term we get
\begin{equation}
 \beta \thav{A}\thav{B} \simeq \beta A(X^*)B(X^*)\exp\left[\frac{1}{2\beta}\left(\frac{\nabla A}{A} \cdot \A^{-1} \cdot \frac{\nabla A}{A}\right)+ \frac{1}{2\beta}\left(\frac{\nabla B}{B} \cdot \B^{-1} \cdot \frac{\nabla B}{B}\right) \right] \sqrt{\frac{\det\H}{\det\A}}\sqrt{\frac{\det\H}{\det\B}},
\end{equation}
with the definition
\begin{equation}
\B \to \B_{i\alpha j\beta} \equiv \H_{i\alpha j\beta} -\frac{1}{\beta}\frac{\partial^2 \log B}{\partial x_{i\alpha}\partial x_{j\beta}}.
\end{equation}
We now expand the exponential in both expressions. Since both are multiplied by $\beta$, we have to keep only the zeroth and the first orders, as all other terms will go to zero in the $\beta\to\infty$ limit. We get
\begin{equation}
 \begin{split}
  \beta[\thav{AB}-\thav{A}\thav{B}] \simeq\ &\beta A(X^*)B(X^*)\left\{\sqrt{\frac{\det\H}{\det\C}} - \sqrt{\frac{\det\H}{\det\A}}\sqrt{\frac{\det\H}{\det\B}}\right.\\
  &\left. + \frac{1}{2\beta}\left[\frac{1}{A(X^*)^2}\nabla A\cdot \C^{-1} \nabla A + \frac{1}{B(X^*)^2}\nabla B\cdot \C^{-1} \nabla B  +\frac{2}{A(X^*)B(X^*)}\nabla A\cdot \C^{-1} \nabla B  \right]\sqrt{\frac{\det\H}{\det\C}}\right.\\
  &\left.-\frac{1}{2\beta}\left[\frac{1}{A(X^*)^2}\nabla A\cdot \A^{-1} \nabla A + \frac{1}{B(X^*)^2}\nabla B\cdot \B^{-1} \nabla B\right]\sqrt{\frac{\det\H}{\det\A}}\sqrt{\frac{\det\H}{\det\B}}\right\}.
 \end{split}
\end{equation}
We must now take the $\beta \to \infty$ limit. The $O(\frac{1}{\beta})$ terms in parentheses are easy to handle, and one gets
\begin{equation}
 [\nabla A \cdot \H^{-1} \cdot \nabla B],
\end{equation}
since
\begin{eqnarray}
 \lim_{\beta\to\infty} \C &=& \H, \\
 \lim_{\beta\to\infty} \A &=& \H, \\
 \lim_{\beta\to\infty} \B &=& \H.
\end{eqnarray}
The zeroth order term requires more caution. At the leading order in $\frac{1}{\beta}$, one has in general
\begin{equation}
 \det(M+\frac{1}{\beta}N) = \det M + \frac{1}{\beta}\det N' + O\left(\frac{1}{\beta^2}\right),
\end{equation}
where $N'$ is a matrix whose first row is the first row of $N$ and all the other rows are the other rows of $M$. This is due to the fact that the determinant of a matrix is a linear application in each of the matrix's rows (or columns). So one gets, for the zeroth order term,
\begin{equation}
\begin{split}
 \sqrt{\frac{\det\H}{\det\C}} - \sqrt{\frac{\det\H}{\det\A}}\sqrt{\frac{\det\H}{\det\B}} =\ &\sqrt{\frac{\det\H}{\det\H - \frac{1}{\beta}\det \C'}} - \sqrt{\frac{\det\H}{\det\H - \frac{1}{\beta}\det\A'}}\sqrt{\frac{\det\H}{\det\H-\frac{1}{\beta}\det\B'}} + O\left(\frac{1}{\beta^2}\right)\\
 =\ &\sqrt{\frac{\det\H}{\det\H - \frac{1}{\beta}(\det \A' + \det\B')}} - \sqrt{\frac{\det\H}{\det\H - \frac{1}{\beta}\det\A'}}\sqrt{\frac{\det\H}{\det\H-\frac{1}{\beta}\det\B'}} + O\left(\frac{1}{\beta^2}\right),
 \end{split}
\end{equation}
and it can now be easily proven that
\begin{equation}
 \lim_{\beta\to\infty} \beta\left(\sqrt{\frac{\det\H}{\det\H - \frac{1}{\beta}(\det \A' + \det\B')}} - \sqrt{\frac{\det\H}{\det\H - \frac{1}{\beta}\det\A'}}\sqrt{\frac{\det\H}{\det\H-\frac{1}{\beta}\det\B'}}\right) = 0.
\end{equation}
So the zeroth order term adds up to zero, and we are left with
\begin{equation}
 \lim_{\beta\to\infty} \beta[\langle A(X)B(X)\rangle - \langle A(X)\rangle\langle B(X)\rangle] = \left.[\nabla A \cdot \H^{-1} \cdot \nabla B]\right|_{X=X^*},
\end{equation}
which is our thesis. In the case $A(X) = B(X) = \frac{1}{V}\frac{\partial U}{\partial\gamma}$, one gets back the expression
\begin{equation}
\frac{1}{V^2} \Xi\cdot\H^{-1}\cdot\Xi
\end{equation}
where $\Xi \equiv \nabla\frac{\partial U}{\partial\gamma}$. We thus recover the know athermal expression~\cite{LemaitreMaloney06,KarmakarLerner10,99ML} for the shear modulus
\begin{equation}
 \mu = \frac{1}{V}\thav{\frac{\partial^2U}{\partial\gamma^2}} - \beta V[\thav{\sigma^2}-\thav{\sigma}^2] \overset{\beta\to\infty}{\longrightarrow} \mu_{Born} - \frac{\Xi\cdot\H^{-1}\cdot\Xi}{V}.
\end{equation}
This shows how, in the thermal case, the mechanism for divergence of the shear moduli as a consequence of the presence of low-lying modes in the Hessian of the potential energy (i.e. marginality in the \emph{mechanical} sense), is now replaced by a mechanism in terms of anomalous fluctuations and, as a result of the fluctuation-dissipation theorem, anomalous non-linear response of the system to external perturbations (i.e. marginality in the \emph{thermodynamic} sense), as discussed in the main text.

\clearpage

\end{widetext}

\bibliography{LJ}

\end{document}